\documentclass[10pt, pra, aps, twocolumn, showpacs, floatfix]{revtex4-1}
\usepackage{graphicx, subfigure,amsmath,amssymb,bm}
\usepackage{color,hyperref,appendix,multirow,braket,physics}
\graphicspath{{images/}}

\newcommand{\lbr}{\left\lbrace}
\newcommand{\rbr}{\right\rbrace}

\begin{document}

\title{Simulated Randomized Benchmarking of a Dynamically Corrected Cross-Resonance Gate}

\author{R.~K.~L.~Colmenar}
\email{ralphkc1@umbc.edu}
\author{Utkan G\"{u}ng\"{o}rd\"{u}}
\author{J.~P.~Kestner}
\affiliation{Department of Physics, University of Maryland Baltimore County, Baltimore, MD 21250, USA}

\begin{abstract}
We theoretically consider a cross-resonance (CR) gate implemented by pulse sequences proposed by Calderon-Vargas \& Kestner, \textit{Phys.~Rev.~Lett.~118, 150502} (2017). These sequences mitigate systematic error to first order, but their effectiveness is limited by one-qubit gate imperfections. Using additional microwave control pulses, it is possible to tune the effective CR Hamiltonian into a regime where these sequences operate optimally. This improves the overall feasibility of these sequences by reducing the one-qubit operations required for error correction. We illustrate this by simulating randomized benchmarking for a system of weakly coupled transmons and show that while this novel pulse sequence does not offer an advantage with the current state of the art in transmons, it does improve the scaling of CR gate infidelity with one-qubit gate infidelity.
\end{abstract}

\maketitle

\section{Introduction}
\label{sec:Intro}

The ability to implement high-fidelity gates is a necessary requirement for creating a fully functional quantum information processor. To this end, fixed-frequency superconducting transmons \cite{Koch_2007} show great promise \cite{Schreier_2008,Houck_2008}, as they have been used to theoretically and experimentally demonstrate one-qubit gates \cite{Barends_2013,Sheldon_2016_1qubit,Motzoi_2009,Chow_2010,McKay_2017} with fidelities as high as 99.97\% \cite{McKay_2017}. However, generating two-qubit entangling operations with similarly high fidelities remains a challenge. A standard approach to entangling fixed-frequency transmons is through the cross-resonance (CR) effect \cite{Rigetti_2010,Chow_2011,Paraoanu_2006,Li_2008}. The CR effect can be observed in a system of two off-resonant fixed-frequency transmons with a small static coupling (e.g., through a quantum bus \cite{Majer_2007}). By irradiating one transmon at the transition frequency of the other, the coupling is modified by a factor whose magnitude is roughly proportional to the ratio of the microwave drive amplitude and the interqubit detuning.

Theoretical considerations have shown that the CR gate is significantly affected by systematic errors attributed to high-energy excitations of the weakly anharmonic transmon and to crosstalk induced by the CR microwave drive \cite{Magesan_2018,Tripathi_2019}. These processes give rise to unwanted terms in the CR effective Hamiltonian. This necessitates the use of control techniques such as composite pulse sequences \cite{Vandersypen_2005,Corcoles_2013} in order to isolate the desired entangling dynamics. In the case of a CR gate, such gate errors can be eliminated by a secondary control pulse on the target qubit which, in conjunction with pulse sequences, can result in CR gate fidelities exceeding 99\% \cite{Sheldon_2016}. However, the pulse sequence used in Ref.~\cite{Sheldon_2016} is not capable of addressing all coherent systematic errors to leading order.

In this paper, we analyze how well a different, recently discovered generic composite pulse sequence \cite{Calderon-Vargas_2017} would perform in the specific application of fixed-frequency transmons coupled via the CR effect as opposed to the conventional approach.  This new sequence inserts local $\pi$ rotations between repeated application of an entangling gate to dynamically correct all coherent systematic errors in that entangling gate, but in practice there is a tradeoff between that reduction of error and the introduction of errors coming from the insertion of imperfect local $\pi$ pulses.  The purpose of this paper is to examine this tradeoff for the case of CR-gated transmons and determine the conditions for which there is a net benefit.

We theoretically simulate standard Clifford randomized benchmarking (RB) to assess the CR gate performance and show that, while there is no benefit to using the sequence of Ref.~\cite{Calderon-Vargas_2017} with current transmon noise levels and single-qubit fidelities, as single-qubit fidelities improve the new pulse sequence could provide better two-qubit RB fidelities than the currently used dynamical correction scheme.

\section{Dynamical Error Correction via Pulse Sequences}
\label{sec:DEC}
We begin by summarizing the formalism developed in Ref.~\cite{Calderon-Vargas_2017}. We are interested in developing a protocol that allows us to dynamically correct coherent systematic error affecting an arbitrary two-qubit entangling gate. To this end, Ref.~\cite{Calderon-Vargas_2017} presented a family of composite pulse sequences that are composed using repetitions of the nonlocal gate
$\left( \theta \right)_{ab} = \exp\left[ - i  \left(\theta/2\right) \sigma_{ab}\right]$, where $a,b \in \lbr X,Y,Z \rbr$, which can be generated from any arbitrary two-qubit coupling along with appropriate one-qubit rotations~\cite{Zhang_2003,Geller_2010}. In practice, the building block $\left( \theta \right)_{ab}$ may contain errors, which we only consider up to the leading order. Thus, we have
\begin{equation}
\label{eq:noisy_U}
\left( \theta \right)_{ab} = \exp\left[ - i  \frac{\theta}{2} \sigma_{ab} \right] \left( I +  i  \sum_{i,j \in \left\lbrace \text{I,X,Y,Z} \right\rbrace} \epsilon_{ij}\sigma_{ij} \right),
\end{equation}
where $\epsilon_{ij}$ is constant in time and is hereafter referred to as the error in the $ij$ error channel. The pulse sequences have the general form
\begin{equation}
\begin{aligned}
\label{eq:general_form}
&\sigma_{\text{echo}}^{\left(n\right)} \left( \theta \right)_{ab} \sigma_{\text{echo}}^{\left(n\right)}
\sigma_{\text{echo}}^{\left(n-1\right)} \left( \theta \right)_{ab} \sigma_{\text{echo}}^{\left(n-1\right)}
\ldots
\sigma_{\text{echo}}^{\left(1\right)} \left( \theta \right)_{ab} \sigma_{\text{echo}}^{\left(1\right)} \\
&= \exp\left[ - i  \frac{\theta}{2} \sum_{l=1}^{n} \xi_l \sigma_{ab} \right] \\
&\times \left\lbrace I +  i  \sum_{i,j} \epsilon_{ij}\sigma_{ij}
\sum_{m=1}^{n}\zeta_{m}^{ij}\exp\left[ i  \frac{\theta}{2} \left( \chi_{ij} - 1\right) \sum_{l=1}^{m-1}\xi_l \sigma_{ab}\right] \right\rbrace,
\end{aligned}
\end{equation}
where $\sigma_{\text{echo}}^{\left(l\right)}$ denotes a local $\pi$ rotation of the form $\sigma_{cd} \equiv \sigma_c \otimes \sigma_d$ with $c, d \in \lbr I, X, Y, Z \rbr$ hereafter referred to as an echo pulse, and
\begin{equation}
\xi_l \equiv \left\{ \begin{aligned} &+1, && \text{if} \left[\sigma_{\text{echo}}^{\left(l\right)},\sigma_{ab}\right] = 0, \\ &-1, && \text{if} \left\{\sigma_{\text{echo}}^{\left(l\right)},\sigma_{ab}\right\} = 0, \end{aligned}\right.
\end{equation}
\begin{equation}
\zeta_{m}^{ij} \equiv \left\{ \begin{aligned} &+1, && \text{if} \left[\sigma_{\text{echo}}^{\left(l\right)},\sigma_{ij}\right] = 0, \\ &-1, && \text{if} \left\{\sigma_{\text{echo}}^{\left(l\right)},\sigma_{ij}\right\} = 0, \end{aligned}\right.
\end{equation}
\begin{equation}
\chi_{ij} \equiv \left\{ \begin{aligned} &+1, && \text{if} \left[\sigma_{ij},\sigma_{ab}\right] = 0, \\ &-1, && \text{if} \left\{\sigma_{ij},\sigma_{ab}\right\} = 0. \end{aligned}\right.
\end{equation}
We refer to a sequence containing $n$ applications of the noisy entangling operation as a ``length-$n$" sequence. To eliminate the effects of the $ij$ error channel to leading order, we require
\begin{equation}
\label{eq:robust}
\sum_{m=1}^{n}\zeta_{m}^{ij}\exp\left[ i  \frac{\theta}{2} \left( \chi_{ij} - 1\right) \sum_{l=1}^{m-1}\xi_l \sigma_{ab}\right] = 0.
\end{equation}
To simplify this robustness condition, Ref.~\cite{Calderon-Vargas_2017} considered two cases: one where only commuting errors are present ($\chi_{ij} = 1$) and one where only anticommuting errors are present ($\chi_{ij} = -1$).

Let us first consider the case where we only have errors that commute with the entangling operation $\left( \theta \right)_{ab}$. In this case, Eq.~\eqref{eq:robust} reduces to
\begin{equation}
\label{eq:robust-l2}
\sum_{m=1}^{n}\zeta_{m}^{ij} = 0.
\end{equation}
This immediately suggests that the robustness constraint is satisfied only for even values of $n$. The robustness condition in Eq.~\eqref{eq:robust-l2} for a length-2 sequence requires $\zeta^{ij}_1 = -\zeta^{ij}_2$. Setting $\zeta^{ij}_1 = 1$ implies that the first echo pulse commutes with all the errors. Without loss of generality, we can choose the first pulse to be the identity operator for simplicity.  Note that, in order to have a non-identity operation, the second pulse must commute with $\sigma_{ab}$, i.e., $\xi_2 = 1$. The second pulse must also anticommute with all the errors in order to satisfy the robustness condition. If every potential commuting errors are present, this is not possible since there is no choice of $\sigma_{\text{echo}}^{(2)}$ that will simultaneously anticommute with all commuting errors, $[\sigma_{\text{echo}}^{(2)}, \sigma_{ij}] = 0 \,  \forall \, ij \, \ni \, [\sigma_{ij},\sigma_{ab}]=0$.  A length-2 sequence can cancel four of the seven commuting error terms while producing an entangling operation, which may be all that is necessary in certain situations, but no more.  (This can be quickly verified for any specific choice of $\sigma_{ab}$ by simply listing all possibilities, but see App.~\ref{app:su2} for the general proof.)

Nonetheless, with the exception of error in the $ab$ channel itself, all errors that commute with $\sigma_{ab}$ can be eliminated to first order by using two nested applications of a length-2 sequence, i.e., a length-4 sequence.  For instance, the length-4 sequence
\begin{align}
&\mathcal{U}^{(4)}\left[\left(\theta\right)_{ab}\right] \equiv \left(\theta\right)_{ab} \sigma_{aI} \left(\theta\right)_{ab} \sigma_{aI} \sigma_{cc} \left(\theta\right)_{ab} \sigma_{aI} \left(\theta\right)_{ab} \sigma_{aI} \sigma_{cc} \nonumber\\
&= \left(\theta\right)_{ab} \sigma_{aI} \left(\theta\right)_{ab} \sigma_{bc} \left(\theta\right)_{ab} \sigma_{aI} \left(\theta\right)_{ab} \sigma_{bc} \nonumber \\
&= \exp\left[ - i  \frac{4\theta}{2} \sigma_{ab} \right]  \left( I + \mathcal{O}\left(\epsilon^{2}\right) \right),
\end{align}
where $\{\sigma_{cc},\sigma_{ab}\}=0$, eliminates all commuting error channels to first order except for the $ab$ channel itself.

We now consider the complementary case where all the errors instead anticommute with the entangling operation $\left( \theta \right)_{ab}$. The robustness constraint in Eq.~\eqref{eq:robust} becomes
\begin{equation}
\sum_{m=1}^{n}\zeta_{m}^{ij}\exp\left[- i  \theta \sum_{l=1}^{m-1}\xi_l \sigma_{ab}\right] = 0.
\end{equation}
Ref.~\cite{Calderon-Vargas_2017} showed that a nontrivial solution can be found when $n = 5$, $\xi_l = 1$, $\zeta_{(1,2,4,5)} = \pm 1$, $\zeta_3 = \mp 1$, and $\theta = \theta_0 \equiv \arccos\left[(\sqrt{13}-1)/4\right] \approx 0.27\pi$. A set of echo pulses that correspond to these values are $\sigma_{\text{echo}}^{(1,2,4,5)} = I$ and $\sigma_{\text{echo}}^{(3)} = \sigma_{ab}$. Thus, a length-5 sequence that corrects all anticommuting errors to leading order is given by
\begin{align}
\label{eq:l5}
\mathcal{U}^{(5)}\left[ \left(\theta_0\right)_{ab} \right] &\equiv \left( \theta_0 \right)_{ab} \left( \theta_0 \right)_{ab} \sigma_{ab} \left( \theta_0 \right)_{ab} \sigma_{ab} \left( \theta_0 \right)_{ab}  \left( \theta_0 \right)_{ab} \nonumber \\
&= \exp\left[ - i  \frac{5\theta_0}{2}\sigma_{ab} \right]\left[ I + \mathcal{O}\left(\epsilon_{anticomm}^2\right) \right].
\end{align}

The resulting gate in Eq.~\eqref{eq:l5} is nearly maximally entangling, but it is not locally equivalent to a $\textsc{cnot}$. We can, however, construct a gate locally equivalent to a $\textsc{cnot}$ that can serve as a two-qubit Clifford group generator by using two applications of the dynamically corrected gate:
\begin{align}
\label{eq:cnot-like}
\mathcal{U}_{\text{Clif}_2} = & \exp\left[- i  \frac{\psi}{2}\sigma^{'}\right] \mathcal{U}^{(5)} \exp\left[- i  \frac{\phi}{2}\sigma^{'}\right] \nonumber \\
& \times \mathcal{U}^{(5)} \exp\left[- i  \frac{\psi}{2}\sigma^{'}\right],
\end{align}
where $\psi = 2 \arctan [ ( \sqrt{-57+16\sqrt{13}}) / ( 4-\sqrt{13} + 2\sqrt{-7+2\sqrt{13}} ) ] \approx 0.36\pi$, $\phi = -2\arccos [ -1/( 2\sqrt{-14+4\sqrt{13}} ) ] \approx -1.56\pi$, and $\sigma^{'} \in \{ \sigma_{IX},\sigma_{IY},\sigma_{IZ},\sigma_{XI},\sigma_{YI},\sigma_{ZI} \}$ such that $\{ \sigma^{'},\sigma_{ab}\} = 0$.

It is possible to combine a length-2 (or length-4) sequence with a length-5 sequence in order to generate a length-10 (or length-20) sequence that can address both commuting and anticommuting error channels simultaneously. Furthermore, all of these pulses can also be combined with a BB1-like pulse sequence in order to correct the $ab$ channel errors. First-order error in this channel can manifest from gate mistiming or fluctuations in the effective interqubit coupling, both of which result in over/under-rotation of the entangling operation. We refer the reader to Ref.~\cite{Calderon-Vargas_2017} for a more detailed discussion.

Finally, we wish to emphasize that although the rest of this manuscript focuses on the application of the length-5 pulse sequence to fixed frequency transmon qubits, similar considerations apply in any other scenario having the key feature that the errors in the entangling gate anticommute with the entangling operator.  For example, in a silicon-based system of two double quantum dots (DQDs), each containing a single spin, coupled through a resonator \cite{Warren_2019}. The resonator is coupled to only one of the quantum dots, which makes the effective coupling dependent on the magnetic gradient within the DQD. Imperfections on the magnetic field gradient, which can be caused by either an anisotropy in the electron g-tensor or misalignment of the local micromagnet, causes systematic commuting and anticommuting errors to emerge. These can be addressed by a length-5 sequence or a combination of a length-2 and a length-5 sequence depending on the severity of the error.  However, from this point on we use numbers appropriate for the CR gated transmon case.

\section{Dynamically Corrected CR Gate}
\label{sec:CR}

\begin{figure*}
\centering
\begin{subfigure}
\centering
\includegraphics[]{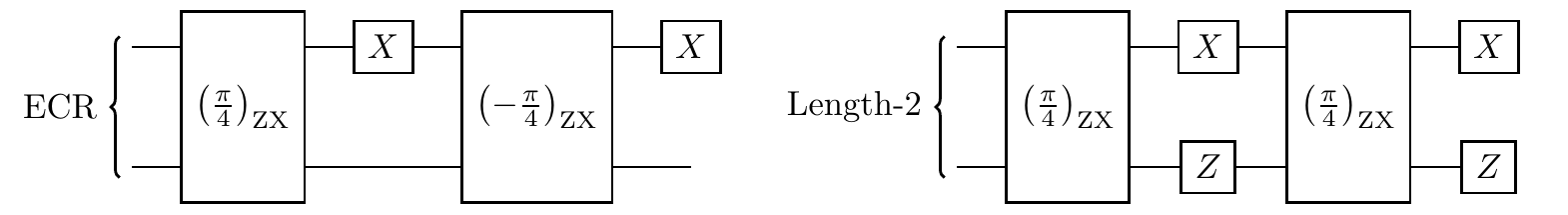}
\end{subfigure}
\begin{subfigure}
\centering
\includegraphics[width=\textwidth,height=\textheight,keepaspectratio]{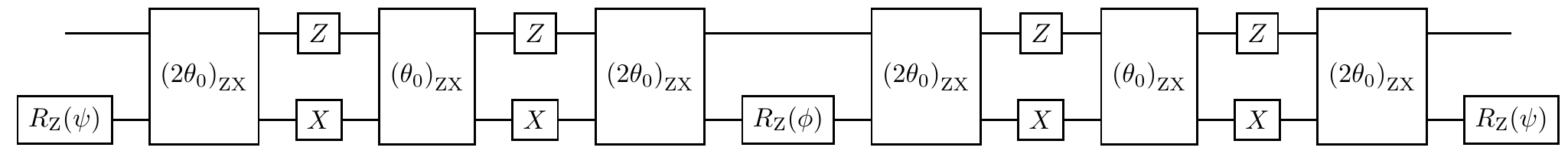}
\end{subfigure}
\newline
\centering
\text{Length-5}
\caption{Circuit diagrams for the two-qubit Clifford generator implemented using the ECR, the length-2 sequence, and the length-5 sequence (see Eq.~\eqref{eq:cnot-like}). The top (bottom) line corresponds to the control (target) qubit. The $X$ and $Z$ gates are the usual Pauli gates and $R_{\text{Z}}(\theta)$ is a rotation about the $Z$-axis by an angle $\theta$. The generated Clifford gate is locally equivalent to a CNOT gate in all cases.}
\label{fig:circuit}
\end{figure*}

We now apply the formalism we summarized in Sec. \ref{sec:DEC} to a CR gate. We consider a system of two fixed off-resonant transmons that are weakly coupled to a bus resonator. We then apply a constant-amplitude microwave driving field on one qubit, the control qubit, at the transition frequency of the other qubit, the target qubit. In the weak driving limit, a block-diagonal effective Hamiltonian for a CR gate can be perturbatively constructed using the Schrieffer-Wolff transformation \cite{Magesan_2018}:
\begin{equation}
\label{eq:ham}
H_{\text{eff}}^{\text{CR}} =  \frac{1}{2}h_{ZI}\sigma_{ZI} + \sum_{j \in \lbr X,Y,Z \rbr} \left(\frac{1}{2}h_{Ij}\sigma_{Ij} + \frac{1}{2}h_{Zj}\sigma_{Zj}\right),
\end{equation}
where the expressions for $h_{ij}$ in terms of the physical parameters are given in the appendix of Ref.~\cite{Magesan_2018}. This approach differs from previous derivation of the CR Hamiltonian \cite{Rigetti_2010} in that it yields coherent error terms pertaining to higher-energy level leakage. We note that the Hamiltonian belongs in the embedding $\mathfrak{su}(2)\oplus\mathfrak{su}(2)\oplus\mathfrak{u}(1) \subset \mathfrak{su}(4)$. In particular, $\mathfrak{u}(1)$ is generated by $\sigma_{ZI}$ which yields a factor in the time-evolution operator that can be removed by applying a local $Z$-rotation on the first qubit. For this reason, we will ignore the effects of the $\sigma_{ZI}$ term. The entangling term here is the $\sigma_{ZX}$ term which, if factored out, yields
\begin{equation}
\label{eq:tevo}
\begin{aligned}
&U(t) = \\
&\exp \left[ -\frac{ i }{2} t h_{ZX} \sigma_{ZX}\right] \left[ I +  i \, \sum_{j \in \lbr X, Y, Z\rbr} \left(\epsilon_{Ij}\sigma_{Ij} + \epsilon_{Zj}\sigma_{Zj}\right)  \right],
\end{aligned}
\end{equation}
where $\epsilon_{ij}$ can be calculated analytically up to a desired order using the Baker-Campbell-Hausdorff (BCH) formula. The pulse sequence building block, $(\theta)_{\text{ZX}}$, can then be obtained by setting $t = \theta / {h_\text{ZX}}$.

In experiments, the microwave drive acting on the control qubit often leaks into the target qubit which results in on-resonant crosstalk. This introduces large $IX$ and $IY$ terms in the effective Hamiltonian. Thus, in practice, the commuting errors are in the $ZX$ and $IX$ channels, while the anticommuting ones are in the $IY$, $IZ$, $ZY$, and $ZZ$ channels \cite{Sheldon_2016}. Neglecting the $ZX$ error channel for the moment, which is reasonable provided that the errors are static and the evolution time of the entangling gate is properly compensated through calibration, the result in Sec.~\ref{sec:DEC} suggests that we apply a length-2 sequence with a $\sigma_{XZ}$ echo pulse to eliminate the $IX$ channel. In addition, this choice of echo pulse can also eliminate any higher-order $ZI$ channel errors that may accumulate due to the presence of large anticommuting error terms. Although alternatives such as $\sigma_{XY}$ can serve the same purpose, we choose $\sigma_{XZ}$ in order to take advantage of novel control methods that allow implementation of near-perfect virtual $Z$-gates via abrupt phase modulation of the microwave control drive \cite{McKay_2017}.

We note that Refs.~\cite{Sheldon_2016} and \cite{Corcoles_2013} use an operationally distinct pulse sequence called an echoed CR (ECR) gate which has the same effect as the above length-2 sequence with an $XZ$ echo pulse. The key operational difference between the ECR scheme and our length-2 sequence is the sign reversal in the entangling operation,
\begin{equation}
\label{eq:ECR}
ECR \equiv \left( \frac{\pi}{4} \right)_{ZX} \sigma_{XI} \left( -\frac{\pi}{4} \right)_{ZX} \sigma_{XI},
\end{equation}
which can be implemented experimentally by reversing the signal of the microwave drive ($\Omega \rightarrow -\Omega$). Unlike in our length-2 scheme, a $\sigma_{XI}$ echo pulse, which anticommutes with $\sigma_{ZX}$, is applied to avoid implementing a purely local gate. We show in App.~\ref{app:ECR} that this yields a mathematically equivalent pulse as the length-2 sequence in the case of a CR Hamiltonian. So in the remainder of our discussions, the length-2 sequence and the ECR gate are equivalent.

Another approach involves applying a secondary microwave pulse onto the target qubit so as to eliminate particular terms in the Hamiltonian \cite{Sheldon_2016}. This cancellation pulse is calibrated such that it eliminates the $h_{IX}$, $h_{IY}$, and $h_{ZY}$ terms. Using the experimental parameters provided in Ref.~\cite{Sheldon_2016}, it can be verified numerically that the remaining terms have different scales, $h_{IZ} \ll h_{ZZ} \ll h_{ZX}$. Thus, the dominant source of remaining error comes from the $h_{ZZ}$ term of the effective Hamiltonian, which translates to errors in the $ZZ$ and $IY$ channels.  Although a length-2 sequence with an $XZ$ echo pulse (i.e., ECR) can partially suppress these anticommuting errors, one can instead get complete first-order correction using the length-5 sequence of Eq.~\eqref{eq:l5}, $\mathcal{U}^{(5)}[U(\theta_0/h_{ZX})]$, with $ab=ZX$.  Then, to obtain a two-qubit Clifford generator, we use Eq.~\eqref{eq:cnot-like} with $\sigma^{'} = \sigma_{IZ}$, where we again make use of virtual $Z$ gates. This yields an entangling Clifford gate compensated for all relevant coherent systematic errors to leading order.

It is important to keep in mind that the theory we summarized in Sec.~\ref{sec:DEC} assumes that the echo pulses can be implemented perfectly. This is not the case in practice and echo pulse errors can be detrimental to the sequence's efficacy. Even though a longer and theoretically better sequence can be obtained by combining the length-2 and length-5 sequences, the resulting length-10 sequence requires more potentially noisy one-qubit gates to implement. So, depending on the level of one-qubit error, a length-2 or length-5 sequence can be more effective than a length-10 sequence. The supplemental material of Ref.~\cite{Calderon-Vargas_2017} indicates that one-qubit gate errors on the order of, at most, $10^{-5}$ are required in order to build a $\textsc{cnot}$ out of a length-20 sequence with gate error below $10^{-3}$. This may be very difficult to realize in the near future, which is why we focus our discussion to the length-2 and the length-5 sequence. In principle, a $\left(\frac{\pi}{2}\right)_{\text{ZX}}$ gate generated using the length-2 sequence requires 4 one-qubit gates, while $\mathcal{U}_{\text{Cliff}_{2}}$ requires 14 one-qubit gates. However, by taking advantage of virtual Z-gates, we can reduce this to 2 and 4 physical one-qubit gates, respectively.

Finally, we note that the two-qubit Clifford gate generated by the length-2 sequence, $\left(\frac{\pi}{2}\right)_{\text{ZX}}$, and by the length-5 sequence, $U_{\text{Clif}_2}$, are different up to local Clifford rotations. In both cases, though, the local invariants of the resulting composite gate are equal to that of a CNOT gate. We present in Fig.~\ref{fig:circuit} a circuit diagram for each of the cases we discussed.

\begin{figure}
\begin{flushleft}
(a)
\end{flushleft}
\includegraphics[scale=.42]{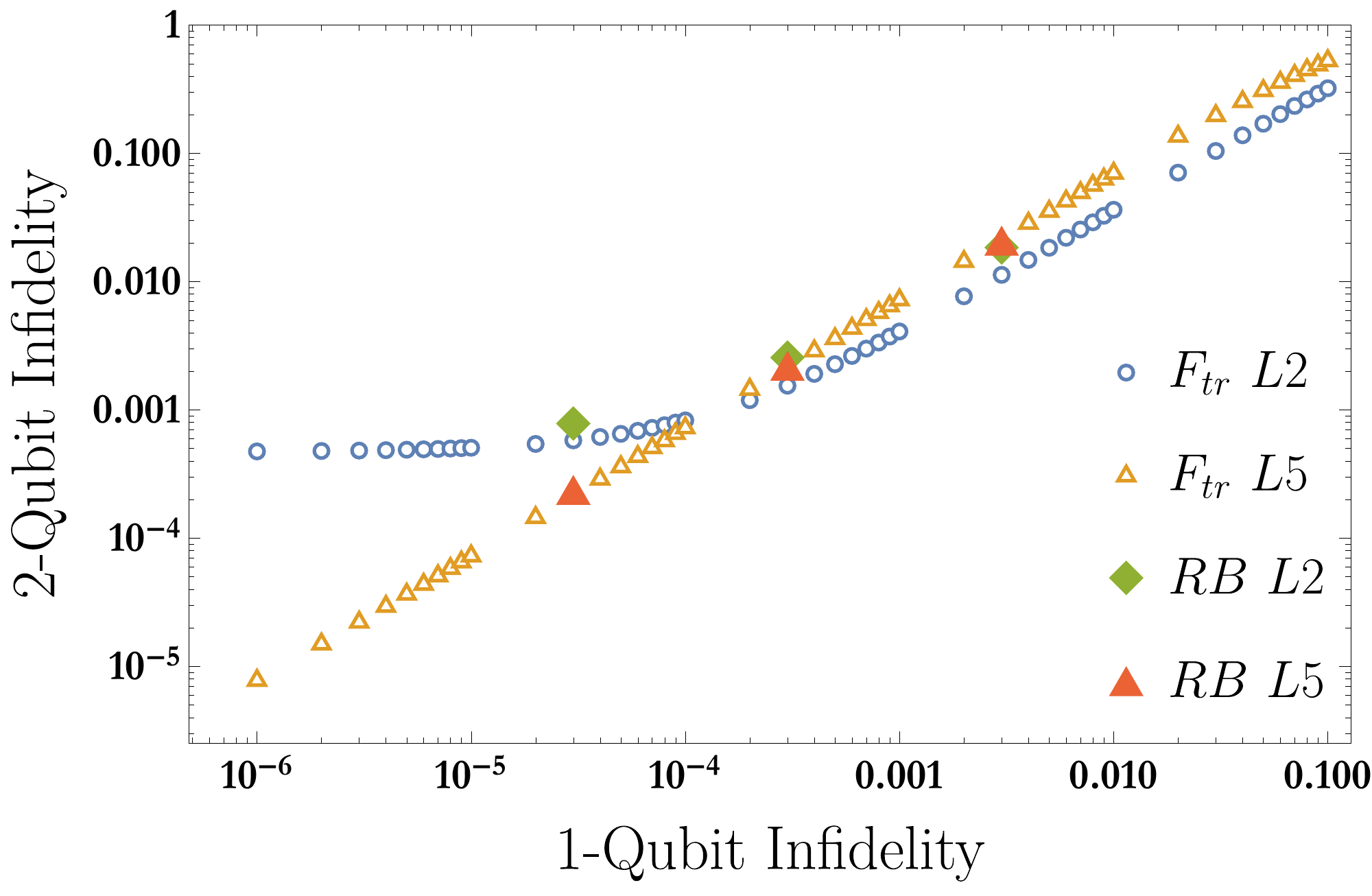}
\begin{flushleft}
(b)
\end{flushleft}
\includegraphics[scale=.42]{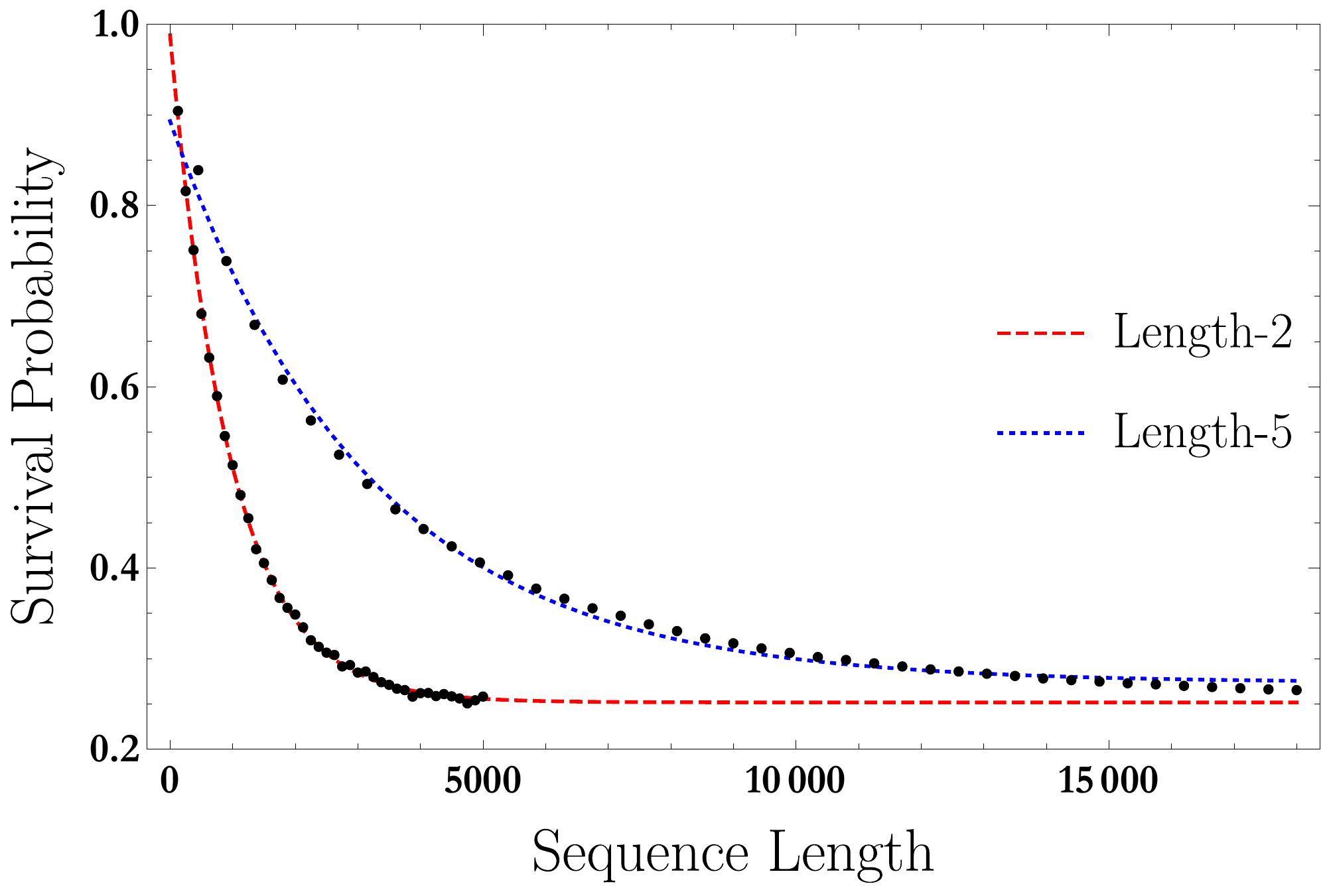}
\caption{(a) Solid symbols show randomized benchmarking (RB) two-qubit Clifford gate infidelity as a function of one-qubit RB infidelity for the length-2 and length-5 sequences. A current experimentally attainable value of one-qubit infidelity is $3\times 10^{-4}$ \cite{McKay_2017}, corresponding to the RB points in the center of the plot. Open symbols show the trace fidelity. (b) A standard RB decay plot comparing the length-2 and length-5 sequences for the case where one-qubit infidelity is set to $3\times 10^{-5}$.}
\label{fig:plot}
\end{figure}

\section{Simulated Randomized Benchmarking}
\label{sec:RB}
To assess the performance of our dynamically corrected gate, we simulate standard Clifford randomized benchmarking (RB) \cite{Magesan_2012} using $\lbr I, X_{\pm \frac{\pi}{2}}, X_{\pm\pi}, Z_{\pm \frac{\pi}{2}}, Z_{\pm\pi}, \mathcal{U}_{\text{Clif}_2}\rbr$ as our generating set where, as an example, $X_\pi$ denotes a $\pi$ rotation about the $X$-axis. We include quasistatic error in all local $X$-rotations using the following noise model:
\begin{equation}
\label{eq:error-model}
X_{\theta} \rightarrow \exp\left[- i  \frac{\varepsilon}{2} \frac{r_x \sigma_X + r_y\sigma_Y + r_z\sigma_Z}{\sqrt{r_x^2+r_y^2+r_z^2}}\right]  X_{\theta},
\end{equation}
where $\lbr r_x, r_y, r_z\rbr$ are sampled uniformly from $\left[-1,1\right]$, and $\varepsilon$ is sampled from a normal distribution centered at $0$ with standard deviation $\delta\theta$. We present in App.~\ref{app:RB} an analytical formula that relates the one-qubit RB infidelity to $\delta\theta$.  On the other hand, $Z$-rotations are performed with no error, corresponding to the virtual gate method described in Ref.~\cite{McKay_2017}.  We also calculate the trace infidelity, which is not efficiently accessible in experiment, but is a less computationally demanding measure for theory, especially in the limit of very weak noise.

We use the experimental parameters reported in Ref.~\cite{Sheldon_2016}: $\omega_{1}/2\pi = 5.114$ GHz, $\omega_{2}/2\pi = 4.914$ GHz, $\delta_{1}/2\pi = \delta_{2}/2\pi = - 0.330$ GHz, $\Omega/2\pi = 60$ MHz, and $J/2\pi = 3.8$ MHz. The evolution time for the building block of the length-2 sequence, $\left(\frac{\pi}{4}\right)_{\text{ZX}}$, is $t = \pi/(4h_{ZX}) = 49.2$ns, while that of the length-5 sequence, $\left(\theta_0\right)_{\text{ZX}}$, is $t = \theta_0/h_{ZX} = 54$ns. In order to simulate the effect of the cancellation pulse, we only include $h_{IZ}$, $h_{ZX}$, and $h_{ZZ}$ in our effective Hamiltonian. Furthermore, we ignore relaxation errors in our simulations and focus solely on coherent systematic error. Each point in the decay curve of our RB simulations are averaged over 1000 different sequences and noise realizations. The sequence length is set just enough to find a good fit for the survival probability function $a p^k + b$, where $a$, $b$, and $p$ are fitting parameters and $k$ is the sequence length. The results of our simulations are presented in Fig.~\ref{fig:plot}.

Note that we do not include the initial portion of the decay from 100\% down to around 90\% (not plotted) in the fit, since there is non-exponential behavior there, particularly in the case of the length-5 decay.  Non-exponential decay is commonly attributed to gate-dependent errors or low-frequency time-dependent noise \cite{Fogarty_2015}, both of which are present in our simulations. Gate-dependent errors are present because we simulated RB with perfect Z-gates but noisy X-gates, as in experiments. Low-frequency noise effects appear because we perform each individual RB run with a fixed set of randomly generated noisy one-qubit Clifford group, again corresponding to the likely experimental case. We keep generating new sets of noisy Clifford gates until we exhaust all of our RB sequences. This builds statistics consistent with the distribution from which the noisy one-qubit gates are generated. A non-exponential decay is obtained by averaging over this ensemble of RB data.

We find that the length-2 sequence performs similarly to the length-5 sequence when the one-qubit RB error is set to $3\times 10^{-4}$ to match Ref.~\cite{McKay_2017}. The length-2 sequence yields a fidelity of 99.7\% and the length-5 sequence, which takes about five times as long (540ns  $ + 4 t_{1\text{Q}}$ vs 98ns  $ + 2 t_{1\text{Q}}$, where  $t_{1\text{Q}}$ denotes the echo pulse gate time), yields 99.8\% \footnote{The reason our simulated length-2 sequence fidelity is slightly higher than the experimental one reported in Ref.~\cite{Sheldon_2016} is simply because we used the more recent lower one-qubit noise value. If we use the conditions of Ref.~\cite{Sheldon_2016}, our trace fidelity calculation yields an error of roughly $6\times 10^{-4}$ which is consistent with the experimentally observed values.}.  However, if the one-qubit errors were reduced, we see that the length-5 sequence increasingly outperforms the length-2.

We can gain further insight by comparing the performance of the two pulse sequences in the limit where there are no one-qubit errors. For this task we use the trace fidelity since simulated randomized benchmarking requires simulating increasingly long sequences to obtain enough fidelity decay to fit as the one-qubit gate error is reduced. We rearrange Eq.~\eqref{eq:noisy_U} and isolate the error terms:
\begin{align*}
\label{eq:error}
\delta U &= \exp \left[ \frac{ i }{2} t h_{ZX} \sigma_{ZX}\right] U(t) - I\\
& =  i \, \sum_{j \in \lbr I, X, Y, Z\rbr} \epsilon_{Ij}\sigma_{Ij} + \epsilon_{Zj}\sigma_{Zj}.
\end{align*}
We numerically calculate this for both schemes, assuming perfect one-qubit gates, and get
\begin{align*}
\delta U_{L2} &= -2.4\times 10^{-4} I + .015 i  (\sigma_{IY} - \sigma_{ZZ}) \\
&\qquad + 7.5 i \times 10^{-4} (\sigma_{IZ} + \sigma_{ZY}) + 3.5 i \times10^{-4}\sigma_{ZX}\\
\delta U_{L5} &= -2 i \times 10^{-5} \sigma_{IX} - 4.8 i  \times 10^{-4}\sigma_{ZX},
\end{align*}
where we have omitted any error terms with magnitudes below $10^{-5}$. Since the gate infidelity is proportional to $\epsilon_{ij}^2$, we see that the length-5 scheme can reach error rates on the order of $10^{-7}$ at best, while the length-2 scheme can only reach $10^{-4}$. The much lower ideal infidelity of the length-5 sequence is because it cancels all the leading order errors in $U(t)$, whereas the length-2 sequence is not capable of eliminating the anticommuting error channels $IY$ and $ZZ$. Of course, the actual performance of both sequences is highly dependent on the severity of the one-qubit gate imperfections, as is evident in Fig.~\ref{fig:plot}, but one can see there that the trace infidelity of the length-5 sequence keeps decreasing with decreasing single-qubit error while the length-2 sequence plateaus in the $10^{-4}$ region.  Moreover, the crossing point where the length-5 is predicted to outperform the length-2 sequence occurs when the one-qubit infidelity is roughly $1\times 10^{-4}$. This indicates that the length-5 sequence may be experimentally viable in the near future if one-qubit gate fidelities can be brought above 99.99\%.

\begin{figure*}
\centering
\begin{subfigure}
\centering
\includegraphics[scale=.53]{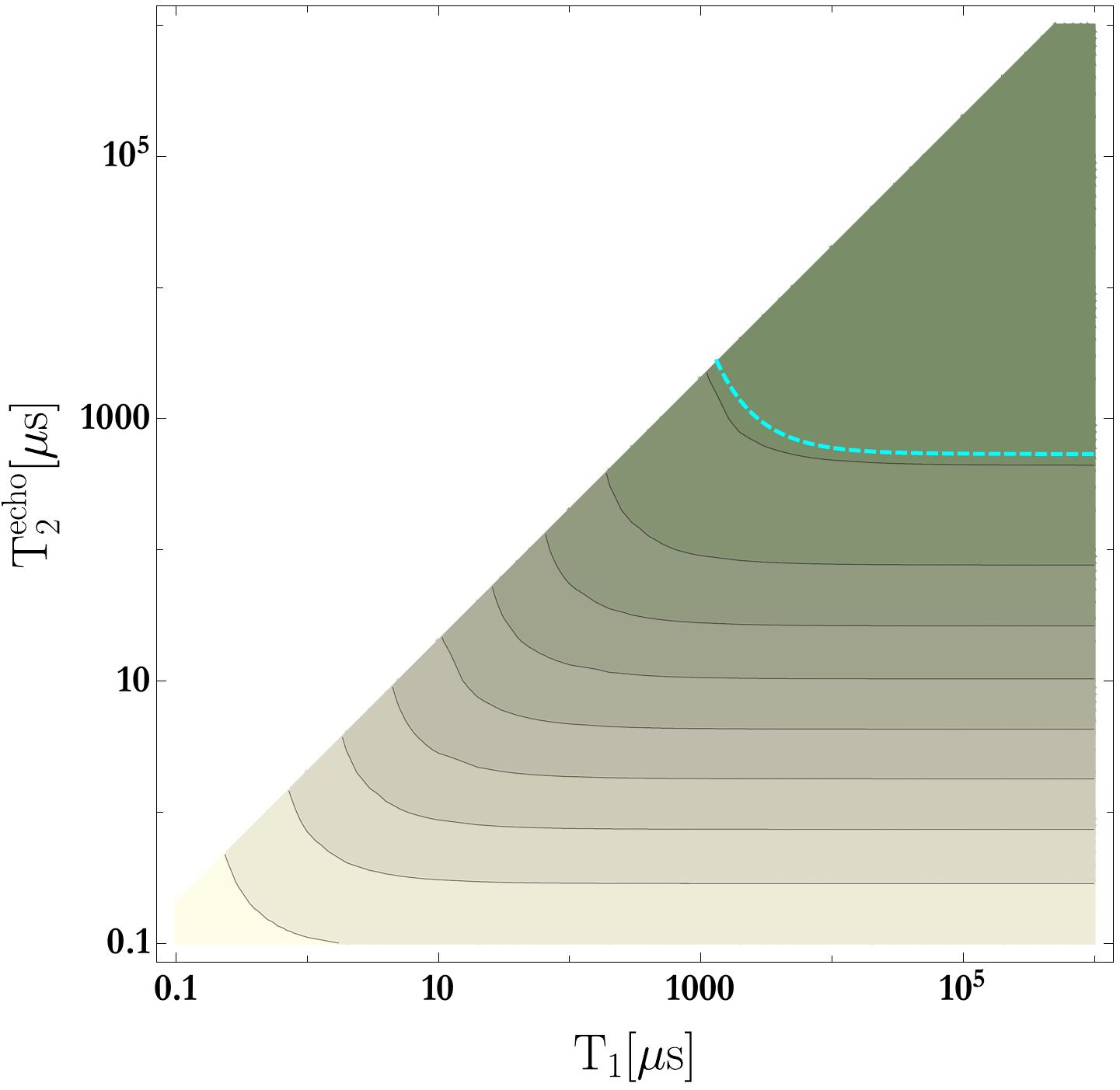}
\end{subfigure}
\begin{subfigure}
\centering
\includegraphics[scale=.53]{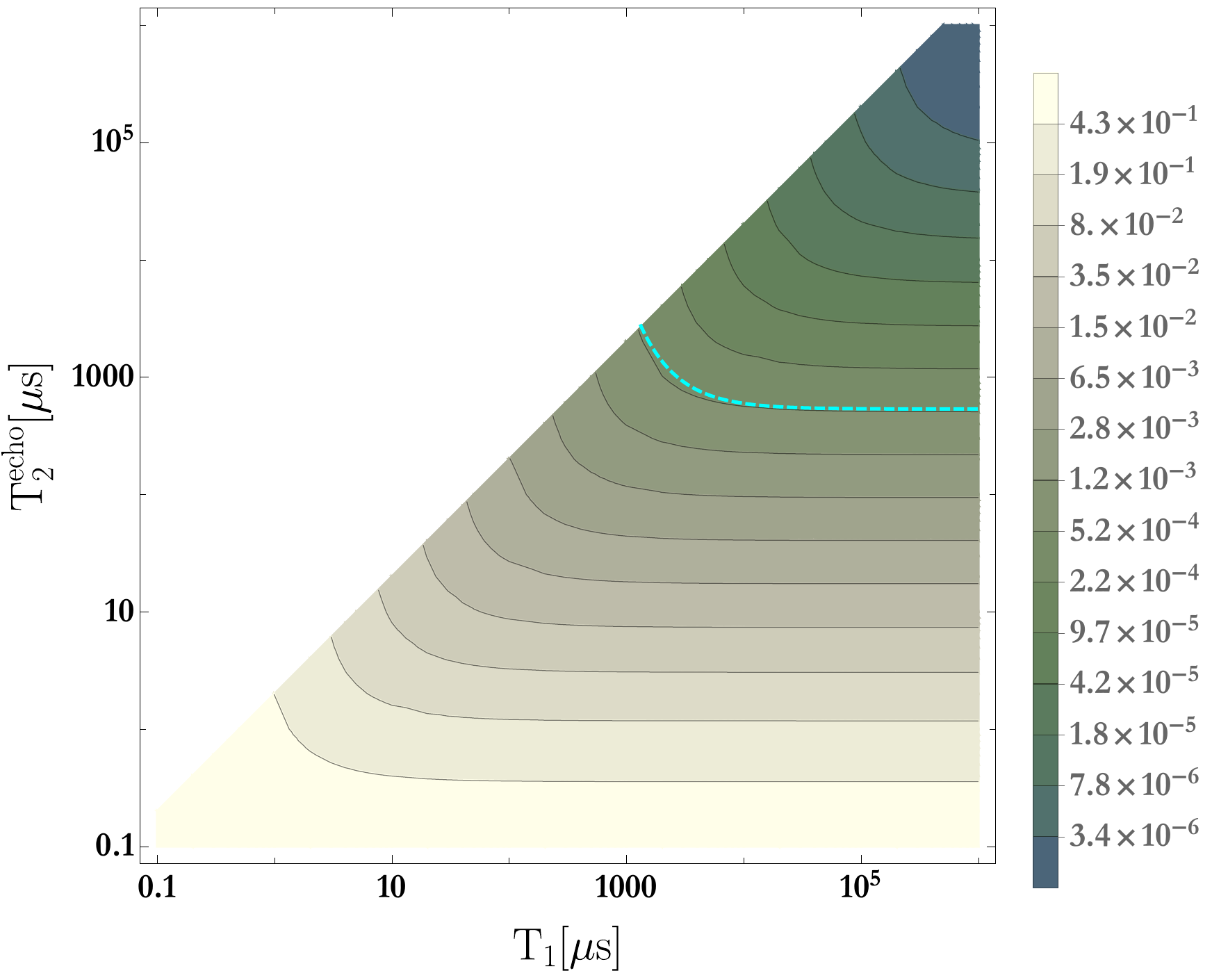}
\end{subfigure}
\caption{A contour plot of two-qubit infidelity as a function of relaxation time, $T_1$, and echoed dephasing time, $T_{2}^{\text{CPMG}}$, for the length-2 sequence (LEFT) and the length-5 sequence (RIGHT), assuming one-qubit gates with no coherent or leakage errors and an average gate time of $30\text{ns}$. Unphysical regions where $T_{2}^{\text{CPMG}} > 2 T_1$ are excluded. The dashed cyan contour indicates the crossing point where the two sequences have equal infidelities. The same colorscale is used for both panels.}
\label{fig:decoherence}
\end{figure*}

One caveat to this conclusion is that, as previously noted, the two-qubit Clifford generated from length-5 sequences is about five times slower than its length-2 counterpart. Thus, the length-5 sequence will suffer more from $T_1$ relaxation error, and its contribution to gate infidelity goes roughly as $T_{\text{gate}}/T_1$ \cite{Chow_2009}. One could consider increasing the CR drive amplitude $\Omega$ to speed up the gate. Numerical analysis of the CR gate in the strong driving regime indicates that the Hamiltonian terms that we considered as systematic error cannot be treated perturbatively when using a na\"{i}ve cosine ramp model for the drive \cite{Tripathi_2019}. These terms can potentially be minimized while reducing the CR gate time by using pulse shapes derived from optimal control schemes which can result in CR gates under 100ns \cite{Kirchoff_2018}. Alternatively, one could also consider increasing the coupling between the qubits to speed up the gate, since the corresponding increase in $\sigma_{ZZ}$ crosstalk due to unwanted excitations to higher energy transmon states would anyways be canceled by the length-5 sequence, but the issue is that the diminished qubit addressability would likely lower \emph{one-qubit} echo pulse fidelities. However, at least the task of engineering a high-fidelity two-qubit gate is then effectively reduced to the problem of engineering high-fidelity local gates.

Without those sort of changes to speed up transmon operations, one has to consider in more detail the trade-off between reduction of coherent error by the length-5 sequence and increased incoherent error due to the longer gate time of the sequence. We aim to quantify this now by analyzing the effects of decoherence. For simplicity, we only consider dephasing and relaxation from the first transmon excited state to the ground state. Using the same parameters as above and setting the ground state energy to zero, we simulate the evolution by a Lindblad master equation
\begin{equation}
\label{eq:Lindblad}
\dot{\rho} = -  i  \left[H_{\text{eff}}^{\text{CR}} , \rho \right] + \frac{1}{T_1} \sum_{j=1,2} \mathcal{D}\left[\sigma_{j}^{-}\right]\rho
+\frac{1}{T_{2}^{\text{CPMG}}} \mathcal{D}\left[\Pi_{j}^{1}\right]\rho,
\end{equation}
where $\rho$ is the density matrix, $T_1$ is the relaxation time of the two qubits, $T_{2}^{\text{CPMG}}$ is the dephasing time measured via Carr-Purcell-Meiboom-Gill (CPMG) pulse sequence (used here as a lower bound on $T_2$), $\sigma_{j}^{-}$ ($\Pi_{j}^{1}$) is the $j^{\text{th}}$ qubit's lowering operator (projection operator to the $\ket{1}$ state), and $\mathcal{D}$ is the damping superoperator
\begin{equation}
\label{eq:damping-supoperator}
\mathcal{D}\left[ A \right]\rho = A \rho A^\dagger - \frac{1}{2} A^\dagger A \rho - \frac{1}{2}\rho A^\dagger A.
\end{equation}
In order to focus on the role of decoherence, we assume that each one-qubit gate in the sequence is implemented without coherent or leakage errors and with an average gate time of 30 ns \cite{Sheldon_2016} during which the transmons can relax. The average two-qubit infidelity can then be calculated \cite{Cabrera_2007}:
\begin{equation}
\expval{F} = \frac{1}{16}\left[ 4 + \frac{1}{5}\sum_{i,j=\text{I,X,Y,Z}} \text{tr}\left[ U \sigma_{ij} U^\dagger \mathcal{M}\left(\sigma_{ij}\right) \right] \right],
\end{equation}
where $U$ is the ideal unitary time-evolution operator, $\mathcal{M}$ is a trace-preserving linear map, and $\sigma_{ij} = \sigma_i \otimes \sigma_j$ are the 15 non-identity Kronecker products of Pauli matrices. We plot the results in Fig.~\ref{fig:decoherence}.

The current state-of-the-art transmons can achieve average coherence times of $T_1 = 0.23\text{ms}$ and $T_{2}^{\text{CPMG}} = 0.38\text{ms}$ \cite{Place_2020}. For those values, as opposed to the case of purely coherent error considered in Fig.~\ref{fig:plot}, Fig.~\ref{fig:decoherence} indicates that even in the absence of one-qubit coherent gate error, the length-5 sequence does not outperform the length-2 sequence due to the effects of incoherent error over the longer gate time. However, at increased coherence times of $T_1 , T_2^{\text{CPMG}} \approx 1.6\text{ms}$, the fidelity of the length-5 sequence begins to surpass that of the length-2 sequence. For $T_1 , T_2^{\text{CPMG}} \gg 1\text{ms}$ the performance of the length-2 sequence plateaus at $3.8\times 10^{-4}$, while the length-5 sequence continues to show improvement until it also eventually plateaus at roughly $3 \times 10^{-7}$, consistent with what we observed in Fig.~\ref{fig:plot}(a).

Thus, while the length-5 sequence is not currently practical, given the rate of improvement in coherence times in recent years (roughly an order of magnitude every three years) \cite{Kjaergaard_2020} and the amount of attention being devoted to this task \cite{Jurcevic_2020}, it is reasonable to expect the length-5 sequence to become a viable option in the near future.

\section{Summary}
\label{sec:Summary}
We have shown how to dynamically correct a CR gate using a recently developed composite pulse sequence and we theoretically simulated a randomized benchmarking protocol for an experimentally accessible comparison of its performance with the standard ECR scheme, which is equivalent to a length-2 pulse sequence. The application of a cancellation pulse onto the target qubit eliminates a significant amount of coherent systematic error from the effective CR Hamiltonian. The length-2 sequence cannot address all of the remaining dominant errors, all of which anticommute with the entangling operation, but the newly developed length-5 sequence can, at the cost of additional local rotations and a slower entangling gate.  We find that both sequences perform similarly against coherent error when using one-qubit gates with currently achievable fidelities. However, we also show that the length-5 sequence performance could scale much better than the length-2 sequence when one-qubit gates are improved.

The pulse sequences we presented can be easily extended to systems with more than two fixed transmon qubits. Ideally, any given pair of control and target qubit must be decoupled from the remaining idle qubits when generating a two-qubit operation. In cases where more than two qubits share the same bus, the static always-on coupling can lead to spurious $Z$ interactions with one or more of the idle qubits. One work-around to this problem is by performing the CR operation on the control and target qubit while decoupling the idle qubits through Hahn-echo-like pulses \cite{Takita_2016,Takita_2017}. We can apply the same idea to the length-2 and length-5 sequence in order to simultaneously address entangling gate errors within the control-target subspace and spurious errors with the idle qubits. However, the additional echo pulses required to implement this makes the sequence even longer than it already is.

The long gate time of the length-5 sequence already makes it impractical for current coherence times, as the improvement the sequence is designed to produce against coherent errors is outweighed by the increased susceptibility to incoherent errors.  However, once coherence times are increased beyond $1\text{ms}$, the sequence we have presented in this paper will become useful for increasing overall two-qubit gate fidelity.

RKLC acknowledges support from the National Science Foundation under Grant Nos.~1620740 and 1915064, and UG from the Army Research Office (ARO) under Grant No.~W911NF-17-1-0287.

\appendix
\newpage
\section{Analysis of the Length-2 Sequence}\label{app:su2}
In Sec. \ref{sec:DEC} we noted that if all potential commuting errors are present then there exists no $\sigma_{\text{echo}} = \sigma_{cd}$ with $c, d \in \lbr I, X, Y, Z \rbr$ that can satisfy
\begin{equation}
\label{eq:l2-limitation}
\{\sigma_{\text{echo}}, \sigma_{ij}\} = 0 \, \forall \, ij \, \ni \, [\sigma_{ij},\sigma_{ab}]=0.
\end{equation}
To prove this, we begin by making the observation that there exists a maximal embedding $\mathfrak{a} \oplus \mathfrak{b} \subset \mathfrak{su}(4)$ \cite{Slansky_1981}, where '$\oplus$' implies commutation of elements between the respective subalgebras. Let us take $\mathfrak{a} = \mathfrak{u}(1)$ with $\sigma_{ab}$ as its generator. This means all the generators of $\mathfrak{b}$, which we denote as $\boldsymbol b = \lbrace b_1, b_2, \ldots b_n \rbrace$, commute with $\sigma_{ab}$. Thus, not only do error channels that commute with $\sigma_{ab}$ belong in this embedding, but the echo pulse also must belong to the corresponding group embedding since it also needs to commute with $\sigma_{ab}$ in order for the sequence to produce a non-identity operation. Without losing any generality, we can partition $\boldsymbol{b}$ into two subsets as
\[
\boldsymbol{b} = \lbrace \overbrace{b_1, b_2, \ldots}^{\{\sigma_{ij}\}} | \overbrace{\ldots b_n}^{\boldsymbol{b}\setminus \{ \sigma_{ij}\}}\rbrace ,
\]
where the left partition contains all the commuting errors that are relevant in the system while the right partition is the coset containing the remaining elements of the generating set of the subalgebra. A $\sigma_{\text{echo}}$ which anticommutes with everything on the left partition and commutes with $\sigma_{ab}$ can only exist in the coset. If all the possible commuting errors are present, then all the elements of $\boldsymbol{b}$ must belong to the left partition, which leaves no possibility for $\sigma_{\text{echo}}$. In other words, if all commuting errors are present, then there exists no $\sigma_{\text{echo}}$ that can satisfy Eq. \eqref{eq:l2-limitation} which proves our claim.

When $\mathfrak{b}$ is a semi-simple Lie-algebra, not all the elements of the coset $\boldsymbol{b}\setminus \{ \sigma_{ij} \}$ will necessarily anticommute with all elements of $\{ \sigma_{ij} \}$. Nonetheless, if the coset happens to contain a $\sigma_{\text{echo}}$ which anticommutes with all the relevant error channels $\sigma_{ij}$, it can be used for error correction. We now show that this is actually the case for the length-2 sequence. But before proceeding further, we first remark that the construction in Section \ref{sec:DEC} relies on the commutation and anticommutation relations of two-qubit Pauli operators. For this reason, we restrict ourselves to embeddings which contain the subalgebra $\mathfrak{b}$ with spinor representation: $\mathfrak{so}(4) \cong  \mathfrak{su}(2)\oplus\mathfrak{su}(2)$ \cite{Ramond_2010}. Given our choice $\mathfrak{a} = \text{span}(\sigma_{ab})$ as the $\mathfrak{u}(1)$ subalgebra, there are two choices for $\mathfrak{b}$.

In the first case with $a=b$, all the elements of the generating set $\lbrace \sigma_{mI}, \sigma_{Im}, \sigma_{nn}, \sigma_{pp}, \sigma_{mp}, \sigma_{pm}\rbrace$, where $m = a$, $n$, and $p$ are mutually distinct and arranged cyclically (e.g. $m = Z, n = X, p = Y$), commute with $\sigma_{aa}$. We can define the generators of the commuting $\mathfrak{su}(2)$ subalgebras as $\sigma_{\widetilde{X}}^{\pm} \equiv (\sigma_{mI} \pm \sigma_{Im})/2$, $\sigma_{\widetilde{Y}}^{\pm} \equiv (\sigma_{np} \pm \sigma_{pn})/2$, and $\sigma_{\widetilde{Z}}^{\pm} \equiv (\sigma_{pp} \mp \sigma_{nn})/2$.

In the second case with $a\neq b$, all the elements of the generating set $\lbrace \sigma_{mI}, \sigma_{In}, \sigma_{nm}, \sigma_{np}, \sigma_{pm}, \sigma_{pp}\rbrace$ commute with $\sigma_{ab}$, where now we have $m = a$ and $n = b$. The generators can be defined as $\sigma_{\widetilde{X}}^{\pm} \equiv (\sigma_{mI} \pm \sigma_{In})/2$, $\sigma_{\widetilde{Y}}^{\pm} \equiv (\sigma_{np} \mp \sigma_{pm})/2$, and $\sigma_{\widetilde{Z}}^{\pm} \equiv (\sigma_{pp} \pm \sigma_{nm})/2$.

In either case, any error channel or echo pulse lies in the subspace spanned by the ``+" and ``-" generators (e.g., $\sigma_{aI} = \sigma_{\widetilde{X}}^{+} + \sigma_{\widetilde{X}}^{-}$). Therefore, any given echo pulse can only anticommute with errors belonging to a different subspace. As an example, since the $\sigma_{aI}$ echo pulse belong to the subspace spanned by the $\sigma_{\widetilde{X}}$ generators, then only errors that belong in the $\sigma_{\widetilde{Y}}$ and $\sigma_{\widetilde{Z}}$ subspaces can be eliminated by a length-2 sequence. Therefore, if all the present commuting errors belong to at most two subspaces only, then the length-2 sequence is sufficient for fixing the errors to first order. The transmon qubit in the main text falls in this category.

A more precise statement of our initial claim is that the length-2 sequence is not capable of correcting errors from all three subspaces. However, placing the initial length-2 sequence inside another length-2 sequence which uses an echo pulse that anticommutes with the initial one allows us to eliminate errors from all three subspaces simultaneously. If we use $\sigma_{aI}$ for our first sequence's echo pulse, the second echo pulse must be in the $\sigma_{\widetilde{Y}}$ or $\sigma_{\widetilde{Z}}$ subspace in order to satisfy the robustness condition.  Clearly, though, the $\mathfrak{u}(1)$ term can not be corrected by a length-2 sequence since it commutes with every allowable echo.

\section{Equivalence of the Length-2 sequence and the ECR scheme}
\label{app:ECR}
In this section we will show that the ECR scheme is mathematically equivalent to a length-2 sequence with a $\sigma_{XZ}$ echo pulse. We begin by noting that the $h_{IZ}$ and $h_{ZZ}$ terms in the effective Hamiltonian of a CR gate are proportional to $\Omega^2$, whereas the $h_{ZX}$ and $h_{IX}$ are only proportional to $\Omega$. Thus, in the absence of noise, the evolution can be generally expressed as
\begin{equation}
U\left(\Omega , t \right) = \exp\left[ - i  t \left(\Omega^2 \left(a \sigma_{IZ} + b \sigma_{ZZ}\right) + \Omega \left(c \sigma_{ZX} + d \sigma_{IX} \right)\right) \right],
\end{equation}
where $a$, $b$, $c$, and $d$ are given in App. C of Ref. \cite{Magesan_2018}. Using the pulse sequence in Eq. \eqref{eq:ECR}, we have
\[
U(\Omega,t) \sigma_{XI} U(-\Omega,t) \sigma_{XI}.
\]
The change from $\Omega \rightarrow -\Omega$ flips the sign of terms that are linearly proportional to $\Omega$. Furthermore, the two $\sigma_{XI}$ surrounding $U(-\Omega,t)$ flip the sign of terms in the exponential which anticommutes with $\sigma_{XI}$. Thus, the cumulative effect of this sequence is
\begin{align}
& U(\Omega,t) \sigma_{XI} U(-\Omega,t) \sigma_{XI} \nonumber\\
&= \exp\Big[ - i  t \big(\Omega^2 \big(a \sigma_{IZ} + b \sigma_{ZZ}) + \Omega \big(c \sigma_{ZX} + d \sigma_{IX} \big)\big) \Big] \nonumber\\
&\times \exp\Big[ - i  t \big(\Omega^2 \big(a \sigma_{IZ} - b \sigma_{ZZ}\big) - \Omega \big(-c \sigma_{ZX} + d \sigma_{IX} \big)\big) \Big].
\end{align}

On the other hand, a length-2 sequence with a $\sigma_{XZ}$ echo pulse yields
\begin{align}
&U(\Omega,t)\sigma_{XZ}U(\Omega,t)\sigma_{XZ} \nonumber\\
&= \exp\Big[ - i  t \big(\Omega^2 \big(a \sigma_{IZ} + b \sigma_{ZZ}) + \Omega \big(c \sigma_{ZX} + d \sigma_{IX} \big)\big) \Big] \nonumber\\
&\times \exp\Big[ - i  t \big(\Omega^2 \big(a \sigma_{IZ} - b \sigma_{ZZ}\big) + \Omega \big(c \sigma_{ZX} - d \sigma_{IX} \big)\big) \Big],
\end{align}
where now we flip the sign of terms in the second exponential which anticommute with $\sigma_{XZ}$. We see that in either case the final products are exactly equivalent.

\section{Analytical Expression for One-Qubit Clifford RB Fidelity}
\label{app:RB}
We now present an analytical expression for the Clifford RB fidelity of a one-qubit gate under the error model given in Eq. \eqref{eq:error-model}. In summary, the goal of Clifford RB is to provide a simple, robust and scalable method for benchmarking the full set of Clifford gates through randomization. The randomization process, also known as twirling, produces a depolarizing channel whose average fidelity can be modeled and experimentally measured. Since the average fidelity of a quantum operation is invariant under the twirling process \cite{Nielsen_2002,Emerson_2005}, the measured fidelity is representative of the original untwirled operation. For a more detailed discussion of Clifford RB, we refer the reader to Ref. \cite{Magesan_2012}.

The key to creating a depolarizing error channel lies in the fact the uniform probability distribution over the Clifford group, $\mathcal{C}$, comprises a unitary two-design. By definition, this gives the twirling condition
\begin{equation}
\label{eq:twirling}
\frac{1}{|\mathcal{C}|}\sum_{i=1}^{|\mathcal{C}|} \left(\mathcal{C}_i^\dagger \Lambda \mathcal{C}_{i}\right)(\rho) = \int_{U(d)} \left(U^\dagger \Lambda U\right) (\rho) \mathrm{d}U,
\end{equation}
where $\mathcal{C}_i$ are elements of the Clifford group, $\Lambda$ is an arbitrary quantum channel acting on the system, and the integral is taken with respect to the Haar measure on $U(2^n)$ with $n$ being the number of qubits. The integral in Eq. \eqref{eq:twirling} produces a unique depolarizing channel $\Lambda_d$ with the same average fidelity as $\Lambda$ \cite{Nielsen_2002,Emerson_2005}. The depolarizing channel is modeled by
\[
\Lambda_{\mathrm{d}}(\rho) = p \rho + (1-p)\frac{I}{2^n},
\]
whose fidelity (as well as $\Lambda$'s) is given by
\[
\mathcal{F} = p + \frac{1-p}{2^n}.
\]

To estimate the average fidelity of one-qubit under the error model given in Eq. \eqref{eq:error-model}, we simply replace $\Lambda$ accordingly and evaluate the sum:
\begin{align}
&\frac{1}{|\mathcal{C}|}\sum_{i=1}^{|\mathcal{C}|} \left(\mathcal{C}_i^\dagger \exp\left[- i  \frac{\varepsilon}{2} \hat{\mathbf{r}}\cdot\vec{\bm{\sigma}}\right] \mathcal{C}_{i}\right) (\rho) \nonumber \\
&= \frac{1}{|\mathcal{C}|}\sum_{i=1}^{|\mathcal{C}|} \mathcal{C}_i^\dagger \exp\left[- i  \frac{\varepsilon}{2} \hat{\mathbf{r}}\cdot\vec{\bm{\sigma}}\right] \mathcal{C}_{i} \rho \mathcal{C}_{i}^\dagger \exp\left[ i  \frac{\varepsilon}{2} \hat{\mathbf{r}}\cdot\vec{\bm{\sigma}}\right] \mathcal{C}_i \nonumber\\
&= \frac{I}{2} + \frac{1+2\cos\left(\theta\right)}{3} \frac{\hat{\bm{\rho}} \cdot \vec{\bm{\sigma}}}{2}.
\end{align}
Since $\rho = \frac{I+\hat{\bm{\rho}} \cdot \vec{\bm{\sigma}}}{2}$, then we must have
\[
p = \frac{1+2\cos\left(\theta\right)}{3}.
\]
Thus, the average Clifford RB fidelity is
\begin{equation}
\mathcal{F} = \frac{2+\cos\left(\theta\right)}{3}.
\end{equation}

In the main text we noted that we used virtual gates in our simulations. This means that any $Z$-gates in the Clifford group are treated as noiseless gates. Thus, we can approximate the fidelity when using virtual $Z$-gates by appropriately weighting the fidelity of the 24 one-qubit Clifford gates:
\begin{equation}
\mathcal{F}_{\mathrm{VZ}} = \frac{20\overline{\mathcal{F}}+4}{24} = \frac{13+5\cos\left(\theta\right)}{18},
\end{equation}
where we assumed that we had 4 noiseless gates ($I,Z,Z_{\pm \frac{\pi}{2}}$). Assuming a Gaussian noise model with a standard deviation $\delta\theta$, we can average over noise realizations and get
\begin{align}
\overline{\mathcal{F}_{\mathrm{VZ}}} &= \frac{1}{\sqrt{2\pi\delta\theta^2}}\int_{-\infty}^{\infty} \exp\left[\frac{-\theta^2}{2\delta\theta^2}\right] \frac{13+5\cos\left(\theta\right)}{18} \mathrm{d}\theta \nonumber \\
&= \frac{13+5\exp\left[-\frac{\delta\theta^2}{2}\right]}{18}.
\end{align}

\bibliography{RB_Fernando_PRL_v7_ref}
\bibliographystyle{apsrev4-1}

\end{document}